\documentclass[11pt]{article}
\usepackage[margin=1in]{geometry}
\usepackage{graphicx}
\usepackage{epsfig}
\usepackage{cite}
\usepackage{float}
\usepackage{hyperref}

\def\beq{\begin{equation}}
\def\eeq#1{\label{#1}\end{equation}}
\def\eeqn{\end{equation}}

\def\beqa{\begin{eqnarray}}
\def\eeqa#1{\label{#1}\end{eqnarray}}
\def\eeqan{\end{eqnarray}}

\let\bar=\overbar

\def\Dslash{\not{\hbox{\kern-4pt $D$}}}
\def\dslash{\not{\hbox{\kern-2pt $\del$}}}

\def\msb{{\bar{\ssstyle M \kern -1pt S}}}

\def\Title#1{\begin{center} {\Large {\bf #1} } \end{center}}
\def\Author#1{\begin{center} {\normalsize {\sc #1} } \end{center}}
\def\Institution#1{\begin{center} {\normalsize {\it #1} } \end{center}}
\def\Abstract#1{\noindent {\normalsize {\bf Abstract:} {\normalfont #1}}}
\def\Conference{\vspace{4mm}\begin{raggedright} {\normalsize {\it ${}^*$Talk presented at the 2019 Meeting of the Division of Particles and Fields of the American Physical Society (DPF2019), July 29--August 2, 2019, Northeastern University, Boston, C1907293.} } \end{raggedright}\vspace{4mm}}

\def\beq{\begin{equation}}
\def\eeq{\end{equation}}
\def\beqa{\begin{eqnarray}}
\def\eeqa{\end{eqnarray}}

\begin{document}

\Title{Soft-gluon corrections for single top quark production in association with electroweak bosons}

\Author{Matthew Forslund$^*$ and Nikolaos Kidonakis}

\Institution{Department of Physics, Kennesaw State University,\\
Kennesaw, GA 30144, USA}

\Abstract{We present results for higher-order soft-gluon radiative corrections for single top-quark production in association with electroweak bosons, including $t\gamma$ and $tZ$ production via anomalous FCNC couplings. We provide results for the total cross sections and differential distributions at LHC energies. We use $K$-factors to show the significance of the corrections compared to leading order, and we discuss uncertainties and the importance of the results.}

\Conference 

\section{Introduction}
Top-quark production at the Large Hadron Collider (LHC) remains an active area of searches for physics beyond the Standard Model, including searches for anomalous couplings \cite{CMSgutgam,CMSgutZ,ATLASgutZ}. In some new physics models, top quarks can be produced via processes involving flavor-changing neutral currents (FCNC). Two such processes are the production of $t\gamma$ \cite{TY,NKAB,ZLLGZ,FG,TopC,DMWZ,DMZ,GYY} and $tZ$ \cite{TY,NKAB,LZ,AAC,TopC,DMWZ,DMZ} pairs without additional quarks in the final state.

The effective Lagrangian involving anomalous FCNC couplings of a $tq$ pair to electroweak bosons, with $q$ an up or charm quark, is given by
\begin{equation}
\Delta {\cal L}^{eff} =    \frac{1}{ \Lambda } \,
\kappa_{tqA} \, e \, \bar t \, \sigma_{\mu\nu} \, q \, F^{\mu\nu}_A + h.c.,
\label{Langrangian}
\end{equation}
where $\kappa_{tqA}$ is the anomalous coupling, with $A$ a photon or $Z$-boson, $\Lambda$ is an effective scale which is taken to be the top-quark mass, $F^{\mu\nu}_A$ are the appropriate electroweak-boson field tensors, and $\sigma_{\mu \nu}=(i/2)(\gamma_{\mu}\gamma_{\nu}-\gamma_{\nu}\gamma_{\mu})$ 
where $\gamma_{\mu}$ are the Dirac matrices.

To improve experimental limits on these anomalous FCNC couplings, it is important to have a good theoretical prediction, which entails studying higher-order corrections. We calculate higher-order corrections to $gq\rightarrow tZ$ and $gq\rightarrow t\gamma$ arising from soft-gluon radiation. These soft-gluon corrections enhance the leading-order (LO) cross sections, and they dominate (and thus approximate well) the higher-order corrections. The approximate next-to-leading order (aNLO) and approximate next-to-next-to-leading order (aNNLO) corrections from soft-gluon emission were studied for the processes $g q \rightarrow t Z$ and $g q \rightarrow t \gamma$ in \cite{gutZ} and \cite{gutgam}, respectively. We present detailed results for cross sections and differential distributions for both processes at LHC energies through aNNLO, and also some approximate next-to-next-to-next-to-leading order (aNNNLO) results. More details on soft-gluon resummation and top-quark production can be found in the review in Ref. \cite{NKrev}.

\section{Soft-gluon corrections}

For the partonic process $g(p_g)+q(p_q) \rightarrow t(p_t)+A(p_A)$,
where $A$ represents a $Z$-boson or a photon, we define the kinematical variables $s=(p_g+p_q)^2$,
$t=(p_g-p_t)^2$, $u=(p_q-p_t)^2$, and $s_4=s+t+u-m_t^2-m_A^2$,
where $m_A=m_Z$ for the $Z$-boson mass and $m_A=m_{\gamma}=0$ for the photon, and $m_t$ is the top-quark mass. 
Near partonic threshold, i.e. when there is just enough
energy to produce the final $tA$ state, but with the top quark and 
electroweak boson not necessarily at rest,  we have $s_4 \rightarrow 0$.
We also define $t_1=t-m_t^2$, $t_2=t-m_A^2$, $u_1=u-m_t^2$, and $u_2=u-m_A^2$.

We consider the double-differential partonic cross section
$d^2{\hat\sigma^{(n)}}_{gq \rightarrow t A}/(dt \, du)$ at $n$th order.
The LO cross section is 
\beq
\frac{d^2{\hat\sigma^{(0)}}_{gq \rightarrow t A}}{dt \, du}
=F^{\rm LO}_{gq \rightarrow t A} \, \delta(s_4) \, ,
\label{LO}
\eeq
where 
\beqa
F^{\rm LO}_{gq \rightarrow t A}&=&
\frac{e^2 \alpha_s \kappa_{tqA}^2}{6s^3 t_1^2}
\left\{2 m_t^6 -m_t^4(3 m_A^2+4 s+2 t) 
+m_t^2\left[2m_A^4-m_A^2(2s+t)+2(s^2+4st+t^2)\right] \right.
\nonumber \\ &&
{}+2m_A^6-4m_A^4t+m_A^2(s+t)(s+5t)-2t(3s^2+6st+t^2) 
\nonumber \\ && \left. 
{}-\frac{t}{m_t^2}\left[2 m_A^6-2m_A^4(s+t)
+m_A^2(s+t)^2-4 s t(s+t)\right]\right\} \, ,
\eeqa
with $\alpha_s$ the strong coupling.

The structure of the soft-gluon corrections is governed by a soft anomalous dimension, $\Gamma^S$. For next-to-leading-logarithm (NLL) accuracy, we need one-loop results.  The one-loop expression for the soft anomalous dimension in Feynman gauge is given by
\beq
\Gamma^{S\, (1)}_{gq \rightarrow tA}=
C_F \left[\ln\left(\frac{-u_1}{m\sqrt{s}}\right)
-\frac{1}{2}\right] +\frac{C_A}{2} \ln\left(\frac{t_1}{u_1}\right)
\label{tgam1l}
\eeq
where $C_F=(N_c^2-1)/(2N_c)$ and $C_A=N_c$, 
with $N_c=3$ the number of colors.
Two-loop \cite{gutZ,gutgam} and three-loop \cite{NK3l} expressions are also available but they are not required for the NLL accuracy that we use here.

The aNLO soft-gluon corrections for $g q \rightarrow tA$ are
\beq
\frac{d^2{\hat\sigma}^{(1)}_{gq\rightarrow t A}}{dt \, du}
=F^{\rm LO}_{gq \rightarrow t A} 
\frac{\alpha_s(\mu_R^2)}{\pi} \left\{c_3 \left[\frac{\ln(s_4/m_t^2)}{s_4}\right]_+
+c_2 \left[\frac{1}{s_4}\right]_+ +c_1 \delta(s_4)\right\} \, ,
\label{NLOgqtA}
\eeq
where
$c_3=2(C_F+C_A)$,
\beq
c_2=2 C_F \ln\left(\frac{u_1}{t_2}\right)-C_F
+C_A \ln\left(\frac{t_1}{u_1}\right)
+C_A \ln\left(\frac{s m_t^2}{u_2^2}\right)
-(C_F+C_A)\ln\left(\frac{\mu_F^2}{m_t^2}\right) \, , 
\eeq
\beq
c_1=\left[C_F \ln\left(\frac{-t_2}{m_t^2}\right)
+C_A \ln\left(\frac{-u_2}{m_t^2}\right)
-\frac{3}{4}C_F\right]\ln\left(\frac{\mu_F^2}{m_t^2}\right)
-\frac{\beta_0}{4}\ln\left(\frac{\mu_F^2}{\mu_R^2}\right) \, ,
\eeq
$\mu_F$ is the factorization scale, $\mu_R$ is the renormalization scale,  
and $\beta_0=(11C_A-2n_f)/3$ is the lowest-order QCD $\beta$ function, with $n_f=5$ the number of light-quark flavors.

The aNNLO soft-gluon corrections for $g q \rightarrow tA$ are
\beqa
\frac{d^2{\hat\sigma}^{(2)}_{gq\rightarrow t A}}{dt \, du}
&=&F^{\rm LO}_{gq \rightarrow t A} 
\frac{\alpha_s^2(\mu_R^2)}{\pi^2} \left\{
\frac{1}{2}c_3^2 \left[\frac{\ln^3(s_4/m_t^2)}{s_4}\right]_+ 
+\left[\frac{3}{2} c_3 c_2 -\frac{\beta_0}{4}c_3\right]\left[\frac{\ln^2(s_4/m_t^2)}{s_4}\right]_+ \right.
\nonumber \\ && \hspace{-20mm}
{}+\left[c_3 c_1 -(C_F+C_A)^2 \ln^2\left(\frac{\mu_F^2}{m_t^2}\right)
-2(C_F+C_A) c_2 \ln\left(\frac{\mu_F^2}{m_t^2}\right)
+\frac{\beta_0}{4}c_3\ln\left(\frac{\mu_R^2}{m_t^2}\right) \right]
\left[\frac{\ln(s_4/m_t^2)}{s_4}\right]_+
\nonumber \\ && \hspace{-20mm} \left.
{}+(C_F+C_A)\left[-c_1 \ln\left(\frac{\mu_F^2}{m_t^2}\right) 
-\frac{\beta_0}{4}\ln\left(\frac{\mu_F^2}{m_t^2}\right)
\ln\left(\frac{\mu_R^2}{m_t^2}\right) 
+\frac{\beta_0}{8} \ln^2\left(\frac{\mu_F^2}{m_t^2}\right)\right] 
\left[\frac{1}{s_4}\right]_+ \right\} \, .
\label{NNLOgqtA}
\eeqa

\section{$gu\rightarrow tZ$ via anomalous top-quark couplings}

\begin{figure}[htb]
\begin{center}
\includegraphics[width=145mm]{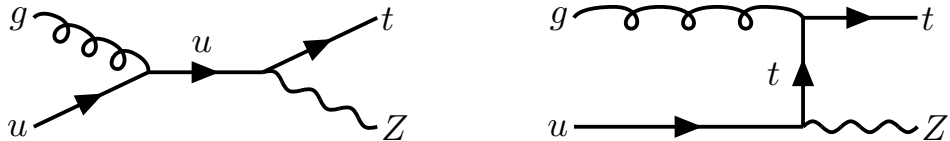}
\caption{Tree-level diagrams for $tZ$ production via anomalous couplings.}
\label{tz}
\end{center}
\end{figure}

\begin{figure}[h]
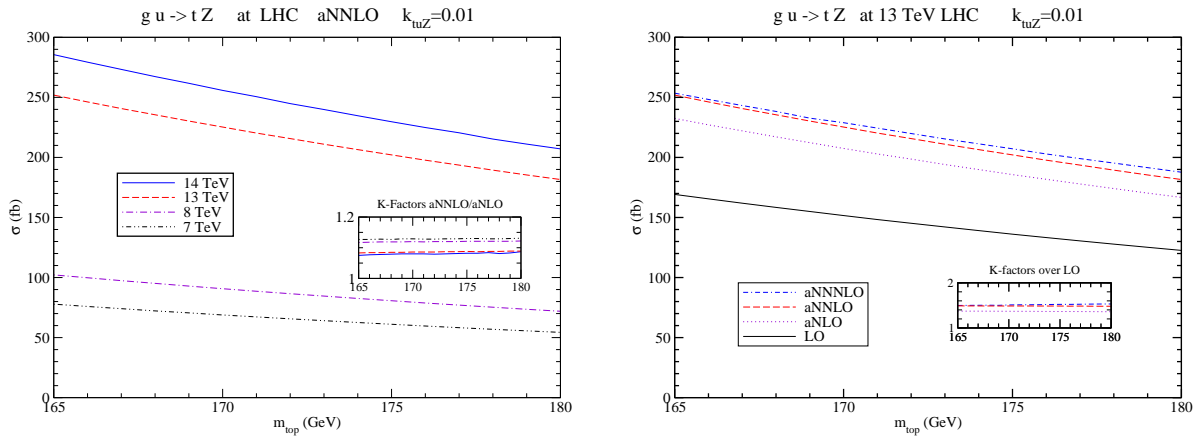

\begin{center}
\includegraphics[width=75mm]{gutzNNLOm.eps}
\hspace{5mm}
\includegraphics[width=75mm]{gutz13mN3.eps}
\caption{The left plot shows results for the total aNNLO cross section for $gu \rightarrow tZ$ at 7, 8, 13, and 14 TeV as a function of top-quark mass. Its inset shows aNNLO/aNLO $K$-factors. The right plot shows the total LO, aNLO, aNNLO, and aNNNLO cross section at 13 TeV as a function of top-quark mass. Its inset shows $K$-factors relative to LO.}
\label{sigmatZ}
\end{center}
\end{figure}

We first present results for $gu\rightarrow tZ$ production. The LO Feynman diagrams contributing to the process are shown in Fig. \ref{tz}. Numerical results for the total cross section are given in Fig. \ref{sigmatZ}. We use MMHT2014\cite{MMHT2014} NNLO parton distribution functions (PDFs) and take $\kappa_{tqZ}=0.01$. The results in \cite{gutZ} use CT14\cite{CT14} PDFs; however, those results are very similar to the ones presented here. The left plot shows the total aNNLO cross section as a function of top quark mass at LHC energies of 7, 8, 13, and 14 TeV, with $K$-factors relative to aNLO shown in the inset. We show a more detailed breakdown at 13 TeV LHC energy in the right plot, showing LO, aNLO, aNNLO, and aNNNLO cross sections, as well as $K$-factors over LO in the inset. We find that the increase in the cross section is substantial at aNNLO for all energies in the plots, with a 49\% increase in the total cross section at 13 TeV at aNNLO compared to the 36\% increase at aNLO. The further contribution from aNNNLO corrections is much smaller.

\begin{figure}[h]
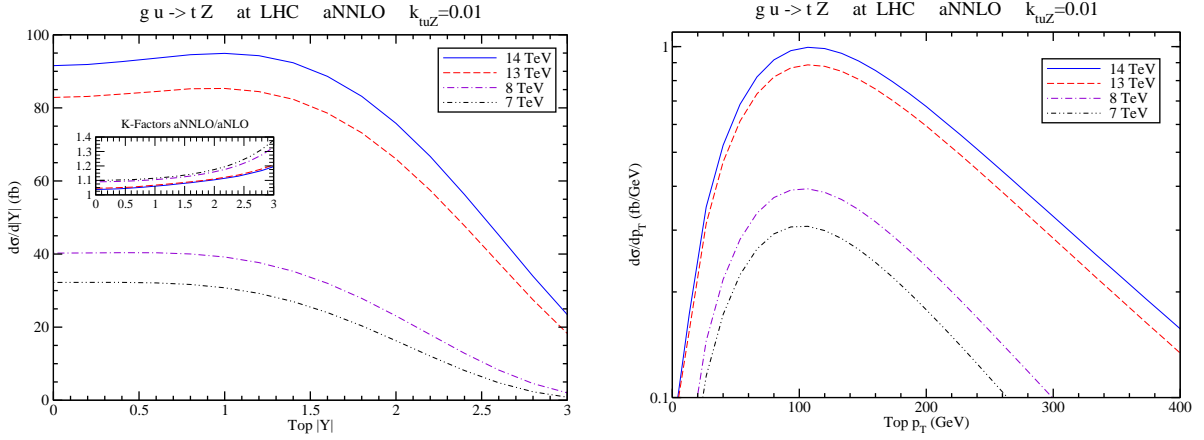

\begin{center}
\includegraphics[width=75mm]{gutzNNLOY.eps}
\hspace{5mm}
\includegraphics[width=75mm]{gutzNNLOpt.eps}
\caption{Differential aNNLO rapidity (left plot) and transverse-momentum (right plot) distributions for $g u\rightarrow tZ$ at LHC energies of 7, 8, 13, and 14 TeV. The inset of the rapidity plot also includes $K$-factors over aNLO.}
\label{tZdiff}
\end{center}
\end{figure}

Differential distributions in top-quark rapidity and transverse momentum are shown in Fig. \ref{tZdiff}, where we see a similarly significant impact by the higher-order corrections. For more details on the soft-gluon corrections to $gq\rightarrow tZ$ and further numerical results, including results for charm-quark initial states and scale dependence, see Ref. \cite{gutZ}.

\section{$gu\rightarrow t\gamma$ via anomalous top-quark couplings}

\begin{figure}[H]
\begin{center}
\includegraphics[width=128mm]{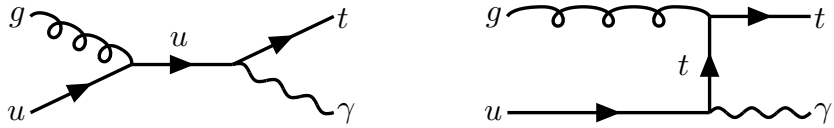}
\caption{Tree-level diagrams for $t\gamma$ production via anomalous couplings.}
\label{tgamma}
\end{center}
\end{figure} 

\begin{figure}[H]
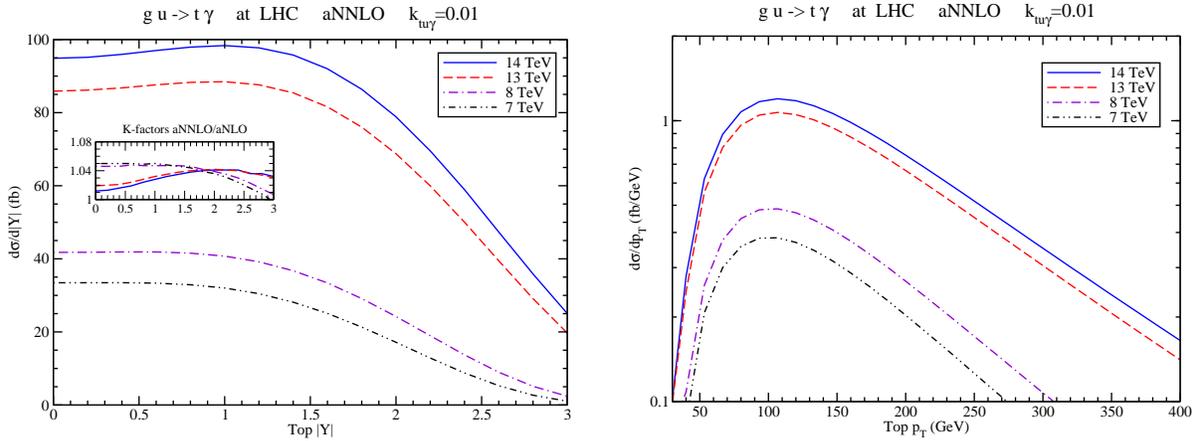

\begin{center}
\includegraphics[width=75mm]{gutgamNNLOY.eps}
\hspace{5mm}
\includegraphics[width=75mm]{gutgamNNLOpt.eps}
\caption{Differential aNNLO rapidity (left plot) and transverse-momentum (right plot) distributions for $g u\rightarrow t\gamma$ at LHC energies of 7, 8, 13, and 14 TeV. The inset of the rapidity plot also includes $K$-factors over aNLO.}
\label{tgamdiff}
\end{center}
\end{figure}

We continue with results for $gu\rightarrow t \gamma$. The LO Feynman diagrams describing the process are shown in Fig. \ref{tgamma}, and are analogous to those for $gu\rightarrow tZ$. Just as with the process $g u \rightarrow t Z$, we use MMHT2014\cite{MMHT2014} NNLO PDFs and take $\kappa_{tq\gamma}=0.01$. The left plot shows results for the total aNNLO cross section as a function of top-quark mass at LHC energies of 7, 8, 13, and 14 TeV, with $K$-factors relative to aNLO shown in the inset. The right plot shows a more detailed breakdown at 13 TeV of the total cross section at LO, aNLO, aNNLO, and aNNNLO, with $K$-factors over LO shown in the inset. Similarly to $gu\rightarrow tZ$, we find a significant increase in the cross section at aNLO and aNNLO, with much smaller contributions from aNNNLO. At 13 TeV, the NLO corrections provide a 31\% increase in the cross section, and the aNNLO results provide a 36\% increase. The differential distributions shown in Fig. \ref{tgamdiff} experience a similarly significant increase by the higher-order corrections. For more details on the soft-gluon corrections to $gq\rightarrow t\gamma$ and further numerical results, including scale dependence and results for charm-quark initial states, see Ref. \cite{gutgam}.

\begin{figure}[H]
\begin{center}
\includegraphics[width=75mm]{gutgamNNLOm.eps}
\hspace{5mm}
\includegraphics[width=75mm]{gutgam13mN3.eps}
\caption{The left plot shows results for the total aNNLO cross section for $g u\rightarrow t\gamma$ at 7, 8, 13, and 14 TeV as a function of top-quark mass. Its inset shows aNNLO/aNLO $K$-factors. The right plot shows the total LO, aNLO, aNNLO, and aNNNLO cross section at 13 TeV as a function of top-quark mass. Its inset shows $K$-factors relative to LO.}
\label{sigmatgam}
\end{center}
\end{figure}

\section{Conclusions}
In some physics models beyond the Standard Model, it is possible to have $tZ$ and $t\gamma$ production without any additional quarks in the final state. In order to have better experimental limits on these anomalous couplings, we have studied higher-order corrections to both the total cross sections and differential distributions for $gq\rightarrow tZ$ and $gq\rightarrow t\gamma$. These corrections dominate the cross section numerically, and approximate well \cite{gutZ,gutgam} the complete corrections at NLO\cite{ZLLGZ,LZ}. At aNNLO, they contribute up to a 36\% increase at 13 TeV for $g u\rightarrow t\gamma$ and 49\% increase at 13 TeV for $g u \rightarrow t Z$. Future work will study soft-gluon corrections in processes with more than two particles in the final state.

\section*{Acknowledgements}
This material is based upon work supported by the National Science Foundation under Grant No. PHY 1820795.

\end{document}